\documentclass[12pt,preprint]{aastex}

\slugcomment{Submitted to ApJ}

\shorttitle{Radio observations of SN 1995N}
\shortauthors{Chandra et al.}

\begin{document}

\title{Eleven years of radio monitoring of the Type IIn supernova SN 1995N}

\author{Poonam Chandra \altaffilmark{1,2},
Christopher J. Stockdale \altaffilmark{3},
Roger A. Chevalier \altaffilmark{2},
Schuyler D. Van Dyk  \altaffilmark{4},
Alak Ray \altaffilmark{5},
Matthew T. Kelley \altaffilmark{6},
Kurt W. Weiler \altaffilmark{7},
Nino Panagia \altaffilmark{8}
Richard A. Sramek \altaffilmark{9}
}

\altaffiltext{1}{Jansky Fellow, National Radio Astronomy Observatory}
\altaffiltext{2}{Department of Astronomy, University of Virginia, P.O. Box
        400325, Charlottesville, VA 22904}
\altaffiltext{3}{Marquette University, Physics Department, P.O. Box
1881, Milwaukee, WI 53214-1881}
\altaffiltext{4}{Spitzer Science Center/Caltech, 220-6, Pasadena, CA 91125}
\altaffiltext{5}{Tata Institute of Fundamental Research, 
Homi Bhabha Road, Mumbai 400005, India}
\altaffiltext{6}{University of Nevada, Las Vegas, Box 454002, Las Vegas,
NV 89154-4002}
\altaffiltext{7}{Naval Research Laboratory, Code 7210, Washington, DC
20375-5351}
\altaffiltext{8}{Space Telescope Science Institute, 3700 San Martin Drive,
Baltimore, MD 21218, USA and INAF/Osservatorio Astrofisico di Catania, Via
S. Sofia 78, I-95123 Catania, Italy and Supernova Ltd., Olde Yard Village
\#131, Northsound Road, Virgin Gorda, British Virgin Islands}
\altaffiltext{9}{National Radio Astronomy Observatory, P.O.~Box 0, Socorro, NM 87801}

\begin{abstract}
We present radio observations of the optically bright Type IIn supernova SN
1995N. We observed the SN at  radio wavelengths with the 
Very Large Array (VLA) for 11 years. 
We also observed it at low radio frequencies with
the Giant Metrewave Radio Telescope (GMRT) at various epochs
within $6.5-10$ years since explosion. 
Although there are indications of an early optically thick phase,
most of the data are in the optically thin regime so it is 
difficult to distinguish 
between synchrotron self absorption (SSA) and free-free absorption (FFA)
mechanisms. However, the information from other wavelengths 
indicates that the FFA is the dominant 
absorption process. Model fits of radio emission 
with the FFA give reasonable physical parameters.
Making use of  X-ray and optical observations, we derive the
physical conditions of the shocked ejecta and the shocked CSM.
\end{abstract}


\section{Introduction}
\label{sec:introduction}

Type IIn supernovae (hereafter SNe IIn) show narrow H$\alpha$
emission line atop a broad emission feature, with no broad absorption.
This classification was  suggested by \citet{sch90} who noticed the above
properties in 8 supernovae (SNe). 
The H$\alpha$ and bolometric luminosities of
most SNe IIn are large, which can be explained by the shock                
interaction of supernova (SN) ejecta with a very
 dense circumstellar (CS) gas \citep{chu90}.
Many SNe IIn also show an excess late time infrared emission,  
which is  a signature of dense CS
dust \citep{ger02}.
CS interaction in SNe is typically observed by radio and X-ray emission
\citep{che03}.
However, in many cases, radio emission from SNe IIn has not been
detected \citep{van96}.

One of the important issues for SNe IIn is
 their high presupernova mass loss rates. For example, for SN 1997ab
 the derived mass loss rate 
for a presupernova
wind  velocity of $90\,{\rm km\, s^{-1}}$ is
$\sim 10^{-2}\, M_\odot\,{\rm yr^{-1}}$ \citep{sal98}. For SN 1994W
the mass loss rate
was $\sim 0.2\, M_\odot\,{\rm yr^{-1}}$ \citep{chu04}, 
and for SN 1995G $\sim 0.1\, M_\odot\,{\rm yr^{-1}}$ \citep{chu03}. 
One possibility for these high mass loss rates is that they are
related to
an explosive event a few years before the SN
outburst \citep{chu03,pas07}. 
Radio observations may put better
constraints on the mass loss of the progenitor stars.

SN 1995N was discovered in MCG $-$02$-$38$-$017 (Arp 261) on 1995 May 5
\citep{pol95} at a distance of 24 Mpc  \citep[see][]{fra02}.
\citet{pol95} estimated that the supernova was at least 10 months old upon
discovery by comparing it with spectroscopic chronometers, and 
classified it as a Type IIn supernova. We
assume the date of explosion to be 1994 July 4 throughout the
paper \citep{fra02}. 
SN 1995N turned out to be relatively bright at late times in most
wavebands. It was detected in the radio
 with the Very Large Array
\citep[VLA,][]{van96} 
and the Giant Metrewave Radio Telescope \citep[GMRT,][]{cha05}.
In X-rays it was detected first by  ROSAT in Aug 1996
and Aug 1997,  and later by  ASCA in Jan 1998
\citep{fox00}. In July 2003 XMM-Newton 
\citep{zam05} and in March 2004  Chandra X-ray
Observatory \citep{cha05} detected the late time X-ray emission from the SN.
SN 1995N has declined very slowly in the optical band \citep{li02,pas05},
with only a 2.5 mag change in the V-band over
$\sim2500$ days after explosion. This  is consistent with the
slow spectral evolution reported in \citet{fra02}.
Ground based optical and HST observations of the late-time spectral
evolution of SN 1995N were used by \citet{fra02} to argue
that the late-time evolution is most likely powered by the X-rays
from the interaction of the ejecta and the circumstellar medium
of the progenitor. They in turn proposed that the progenitors
of Type IIn supernovae are similar to red supergiants in their
superwind phases, when most of their hydrogen-rich gas is expelled
in the last $10^4\, \rm yr$ before explosion.

In this paper, we report extensive radio observations of SN 1995N
taken with the VLA and GMRT. We discuss the observations and data
analysis in \S \ref{sec:observations}. We fit various models to the data and
discuss the results in \S \ref{sec:absorption}. In
\S \ref{sec:interpretation}, we utilize the results from other wavelength observations 
along with the radio bands to derive constraints on the physical parameters.
We summarize our results briefly in \S \ref{sec:conclusions}.        

\section{Observations}
\label{sec:observations}

\subsection{VLA observations}
\label{sec:vla}

The VLA followed SN 1995N for eleven years, from June 1995 until
September 2006. The observations were taken at 22.48 GHz (1.3 cm), 
14.96 GHz (2 cm), 8.46 GHz (3.6 cm), 4.86 GHz (6 cm), 
1.46 GHz (20 cm) and 0.33 GHz (90 cm) frequencies in standard continuum mode
with a total bandwidth  of $2 \times 50$ MHz. 
SN 1995N was observed in all the VLA array configurations.
For the first year of VLA  observations, i.e. until May 1996, 
J1507$-$168 was used as a
phase calibrator. From Oct 1996 until Jan 2004, 
the VLA calibrator
J1504$-$166 was used for phase calibration. 
For only  one epoch, in Sep 2004 at 6 cm, 
J1512$-$090 was used as the phase calibrator. After that until Sep 2006,
J1507$-$168 was used for phase calibration.
Table \ref{tab:phase}
 gives the flux density values of the phase calibrators at each epoch.
 Calibrators 3C48 and 3C286 were 
used for flux calibration at various epochs.
In some cases, where no primary calibrator was observed, we 
obtained those datasets which were having primary calibrators, and 
most closely spaced data in time,
in most cases within a gap of one to two days. 
We used those datasets to perform the
flux calibration on SN 1995N. We use the
Astrophysical Image Processing System (AIPS) 
to analyze the VLA data.
Our A-configuration data on 1996 0ct 24  in 22.5 GHz band was the 
highest resolution data, which led us to derive
the most accurate radio  position of SN 1995N. 
The B1950 position for SN 1995N from this dataset 
is RA: $14^{\rm h}~46^{\rm m}~46.5946\pm0.0004^{\rm s}$, Dec:
$-09^{\rm o}~57'~49.5596\pm0.0006''$.

\subsection{GMRT observations}
\label{sec:gmrt}

We sampled the light curves of SN 1995N at
low frequencies between $6.5-10$ years after explosion using the
GMRT. We observed the SN on around fifteen occasions 
in the 1420 (20 cm), 610 (50 cm), 325 (90 cm) and 235 (125 cm) MHz
bands \citep{cha05b}.
The total time spent at the SN field of view (FOV)
during the observations at various epochs ranged from
2 to 4 hours.  About
17 to 29 good antennas could be used in the radio
interferometric setup at different observing epochs.
For 20 cm, 50 cm and 90 cm observations
the bandwidth  was 16 MHz divided into 128 frequency channels
(the default for the GMRT correlator),
while a 6 MHz bandwidth was used for the 125 cm observations.
3C286 was the main flux calibrator for all the
observations. On a single occasion at 50 cm, we also used 3C147 for flux 
calibration.
In 20 cm observations,
we used J1432$-$180 as the phase calibrator
at most epochs. On a few occasions we used J1347$+$122,
J1445$+$099 and J1351$-$148 for phase calibration.
For the low frequency observations, in the 50 cm, 90 cm and 125 cm bands, we
used J1419$+$064 as phase calibrator at all epochs.
Table \ref{tab:phase}
 gives the flux densities of the phase calibrators at each epoch.
We used the above flux calibrators and
phase calibrators for bandpass calibration as well.
Flux calibrators
were observed once or twice for $10-20$ minutes during each observing session.
Phase calibrators were observed for $5-6$ minutes after every 25 minute
observing run on the SN. 

AIPS was used to analyze all datasets with the standard GMRT data reduction.
Bad antennas and corrupted
data were removed using standard AIPS routines. 
To take care of
the wide field imaging at low GMRT
frequencies, we divided the whole field of view into 5 subfields for the
 1420 MHz frequency datasets, and into $6-9$ subfields for the 610
MHz datasets. For the 325 MHz datasets, wide field imaging
was performed with $16-20$ subfields, while for the 245 MHz datasets
this was done with $25-30$ subfields.
Bandwidth smearing was taken care of by dividing the whole band of
16 MHz into 6
subbands in the 325 MHz observations. In the 235 MHz data analysis, we
divided the 6 MHz band into 6 subbands. For the 1420 MHz and 610 MHz
bands, the bandwidth smearing was insignificant.
A few rounds of phase self-calibrations were also performed on
all the datasets to remove the phase variations due to the weather
and related causes. 

\begin{figure}
\includegraphics[angle=0,width=\textwidth]{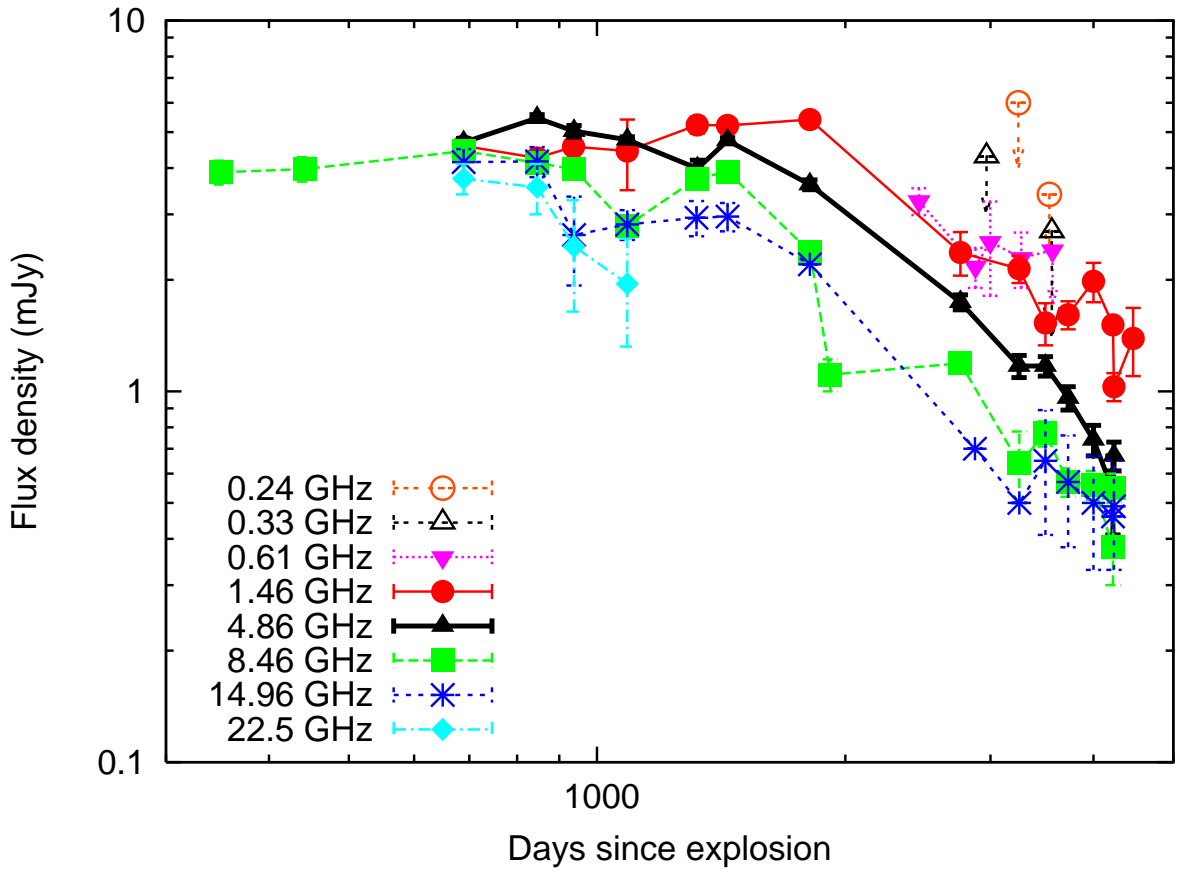}
\caption{Light curves of SN 1995N in 
the 0.24 GHz, 0.33 GHz, 0.61 GHz, 1.46 GHz, 4.86 GHz, 8.46 GHz, 
14.96 GHz and 22.50 GHz bands.} 
\label{fig:full-lc}
\end{figure}

The details of each observation and the SN 1995N flux densities
at various epochs
are tabulated in Table \ref{tab:radio}.
 In Fig. \ref{fig:full-lc} we plot the full  dataset at all
frequencies.

\section{Radio Absorption Process}      
\label{sec:absorption}

The radio emission in a SN is from the shocked interaction shell and is
non-thermal synchrotron in nature. In the simplest models, it can initially be
absorbed either by the internal medium due to synchrotron self 
absorption (SSA) or by the external medium 
through free-free absorption (FFA),
depending on the mass loss rate of the 
presupernova star, magnetic field in the shocked shells, density
of the ejecta and the CS parameters. 
Unfortunately, a clear distinction between
the two absorption processes lies in the optically thick part of 
the light curve or spectrum, where the data for SN 1995N are quite sparse. 
Therefore, before attempting detailed model fits, 
we try to extract information about  both absorption 
processes from other wavebands. 

We can relate the free-free optical depth of the
external medium to the 
hydrogen Balmer line luminosity,
which can be written as \citep{ost89} 
\begin{equation}
L_{\rm H\beta}=4.2\times 10^{-26}\int_R^\infty n_R^2\left(\frac{r}{R}
\right)^{-4} 4 \pi r^2 \, dr.
\label{eq:hbeta}
\end{equation}
Here, cgs units are used and we assume the
 electron temperature of the CSM to be 20,000 K, deduced
by \citet{fra02} using the optical narrow line emission for SN 1995N.
The radius $R$ is the outer radius of the shock and $n_R$ is the 
CSM density at a distance $R$ from the explosion center.
 \citet{fra02} estimate
the SN H$\beta$ luminosity to be $5.3 \times 10^{36}$ erg s$^{-1}$.
Thus, Eq. (\ref{eq:hbeta})  gives $n_R^2 R^3=10^{61}$ cm$^{-3}$.          
The free-free optical depth  for this medium is
$\tau_{\rm FFA}=\int_R^\infty \kappa_{ff} n_R^2 \left(\frac{r}{R}
\right)^{-4} dr$. 
Substituting $n_R^2 R^3=10^{61}$ erg s$^{-1}$, this becomes
\begin{equation}
\tau_{\rm FFA}^{\rm H\beta}=0.5 \left(\frac{\nu}{5\,\rm GHz}\right)^{-2} 
\left(\frac{T}{2 \times 10^4 \rm K}\right)^{-3/2} 
\left(\frac{R}{10^{17}\,\rm cm}\right)^{-2}. 
\end{equation}
Here $\tau_{\rm FFA}^{\rm H\beta}$ implies FFA optical depth 
derived from H$\beta$ luminosity.
This exercise demonstrates that the
optical depth derived from the H$\beta$ luminosity in the 
visual band reaches order unity at radio
wavelengths, indicating that  the
FFA optical depth is sufficient to explain the radio absorption.

\citet[][Fig. 7]{che03} plot the peak radio
luminosity vs the time of peak radio flux for various
supernovae, marking the lines of
constant shock velocity derived assuming SSA absorption process.
If we place SN 1995N in this plot, it falls close to a velocity of
3000 km s$^{-1}$. However, the optical line emission observations
suggest ejecta velocities of at least 5000 km s$^{-1}$ \citep{fra02}.
Hence, the  velocity of the radio-emitting region must be at least
 be equal to or greater than 5,000~km~s$^{-1}$, an indication
 that SSA is not a significant factor in the the radio
 emission measured from SN~1995N.  
 The H$\beta$ luminosity and the inferred expansion 
of the radio-emitting region suggest that FFA
 is the dominant radio absorption mechanism. 

\begin{figure}
\includegraphics[angle=0,width=.79\textwidth]{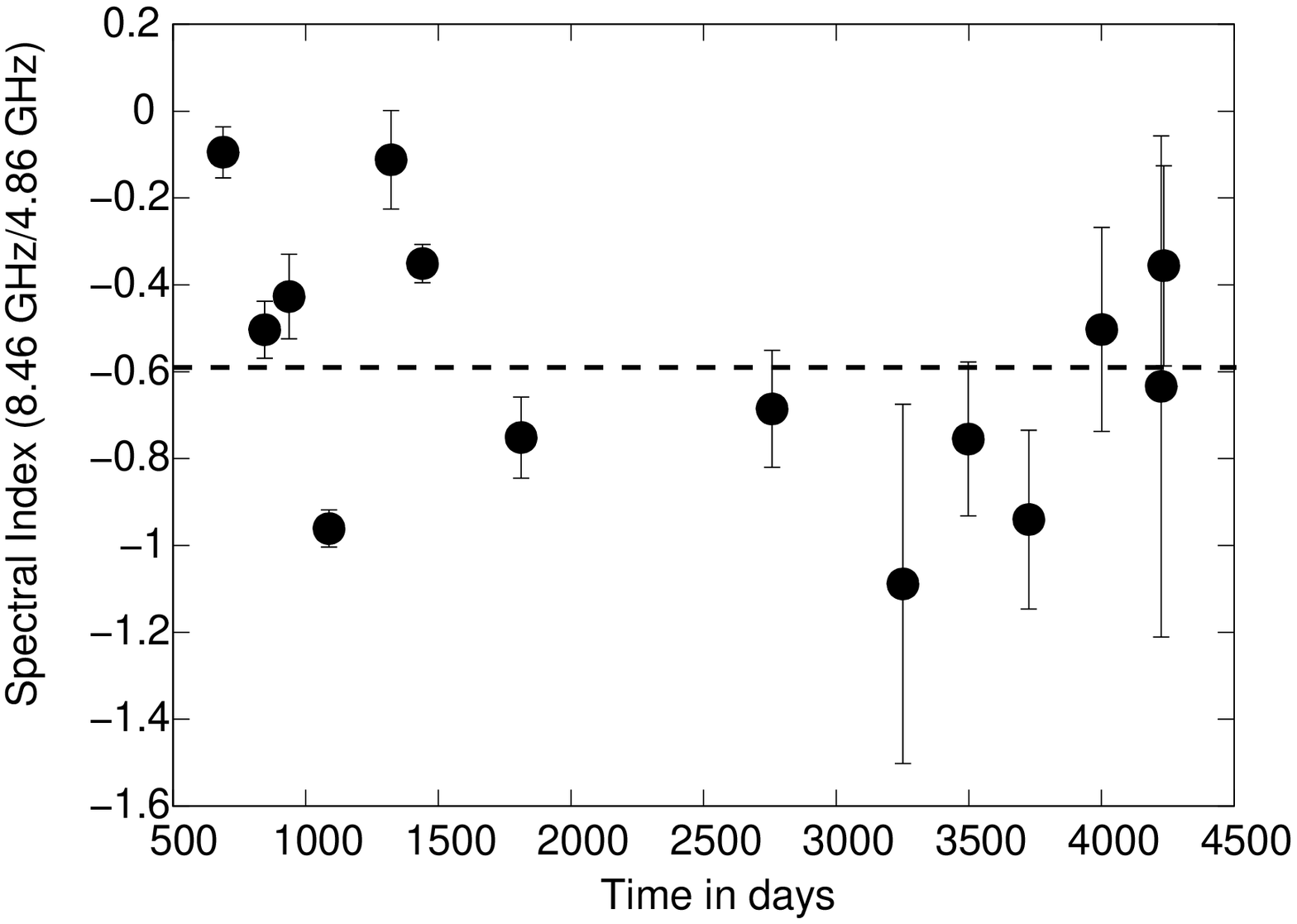}
\includegraphics[angle=0,width=.79\textwidth]{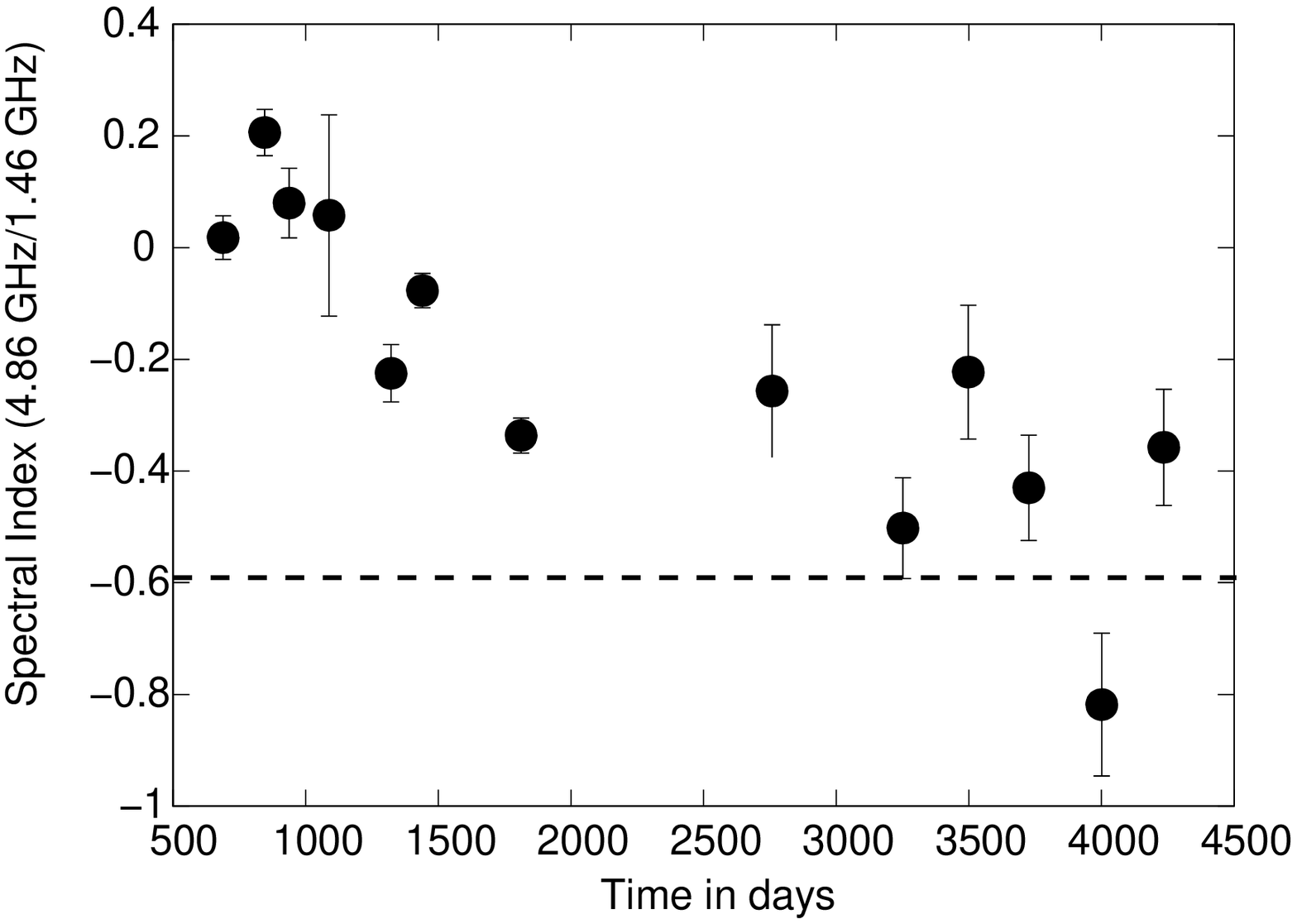}
\caption{Spectral index evolution between 8.46 GHz and 4.86 GHz (upper panel),
and between 4.86 GHz and 1.46 GHz. 
The thick dashed  line is plotted by averaging
all the values of the spectral index after day 2500.}
\label{fig:alpha}
\end{figure}

We now fit the FFA model to the data.            
The model for FFA was given by \citet{che82} and
was formulated in detail by \citet{van94} and \citet{wei02}. The
radio flux density, $F(\nu,t)$, under the FFA 
assumption can be written as:
\begin{eqnarray}
F(\nu,t)=A_1 
\left(\frac{\nu}{5 \,{\rm GHz}}\right)^{-\alpha}
\left(\frac{t}{1000\, {\rm day}}
\right)^{-\beta} 
\exp(-\tau_{\rm FFA}) \nonumber \\ 
\tau_{\rm FFA}=A_2 \left(\frac{\nu}{5 \,{\rm GHz}}\right)^{-2.1}
\left(\frac{t}{1000\, {\rm day}} \right)^{-\delta},
\label{eq:ffa}
\end{eqnarray}
where $\alpha$ is the frequency spectral index, which relates to electron 
energy index $\gamma$ ($N(E) \propto E^{-\gamma}$) as
$\gamma=2 \alpha +1$.  To determine the value of $\alpha$, 
we  plot spectral indices between 8.46 GHz and 4.86 GHz
bands ($\alpha[8.46/4.86]$)
 as well as between 4.86 GHz and 1.46 GHz bands ($\alpha[4.86/1.46]$)
 at various epochs (see Fig \ref{fig:alpha}).  From Fig. 
\ref{fig:full-lc}, it is evident that the light curves becomes optically
thin for these three bands after day $\sim 2500$. Thus we average  
both $\alpha(4.86/1.46)$ and $\alpha(8.46/4.86)$ for all the 
values after day 2500. 
The average spectral index is $\alpha=0.59\pm0.07$. 
Hence, we use $\alpha=0.6$ ($\gamma=2.2$) in the radio absorption models.
Here $A_1$ is the radio flux density normalization 
parameter and $A_2$ is the FFA optical depth normalization parameter.
The parameter $\delta$ is related to the
expansion parameter $m$ in $R \propto t^{m}$ as $\delta=3 m$. 
The assumption that the energy density in the
particles and the fields is proportional to the postshock energy density
leads to $\beta=(\gamma+5-6m)/2$ \citep{che82}. 
For $\gamma=2.2$, this becomes $\beta=3.6-3m$.
Using these values, we fit the SN 1995N radio data to Eq. \ref{eq:ffa}.
The parameters for the best fit model are:
$A_1= 4.59 \pm 0.19$, $A_2=(4.58 \pm 0.80) \times 10^{-2}$, 
and $m=0.84\pm0.02$. 
The above fit parameters yield 
$\delta=2.54\pm0.06$,
and $\beta=1.09\pm0.06$. 
In the CS interaction model of \citet{che82}, the power law index of the supernova
density profile, $n$, is related to $m$ by $m=(n-3)/(n-2)$, so that the above
value of $m$ implies $n=8.3$.

Although our fits for the FFA model are reasonable, we try to fit 
 other radio absorption models as well.
We use the formulation of \citet{che98} for the
radio flux density with SSA present:
\begin{eqnarray}
F(\nu,t)=P_1 \left(\frac{t}{1000\, {\rm day}}\right)^a \left(\frac{\nu}{5\, {\rm GHz}}\right)^{5/2}
 \left(1-\exp(-\tau_{\rm SSA}) \right), \nonumber \\ 
\tau_{\rm SSA}=P_2 \left(\frac{t}{1000\, {\rm day}}\right)^{-(a+b)}
\left(\frac{\nu}{5\, {\rm GHz}}\right)^{-(\gamma+4)/2}.
\label{eq:ssa}
\end{eqnarray}
Here $a$ gives the time evolution of the radio flux density in the
optically thick phase ($F\propto t^a$) 
and $b$ in the optically thin phase ($F\propto t^{-b}$).
Under the assumption that the energy density in the
particles and the fields is proportional to the postshock energy density, 
these quantities are related with expansion parameter $m$ 
and electron energy index $\gamma$ as $a = 2m + 0.5$,   
and  $b=(\gamma+5-6m)/2$ (as before). 
The estimation of
$b$ from the fits is expected to 
be  robust because there are ample data points in
the optically thin part of the light curve, whereas $a$
is determined by the optically thick regime and cannot be constrained
well due to sparse data in this region.
Here $P_1$ is the flux normalization parameter 
and $P_2$ is  the SSA optical depth normalization parameter.
The best fit model gives 
$a=5.97\pm1.13$ and $b=1.12\pm0.07$. 
For $\gamma=2.2$, the above value of $b$ corresponds to $m=0.83$.
The  best fit value of $a$  corresponds to $m\gg 1$. 
However, $a$ is not well constrained by the data, so we fix it to a value of 
$a=2.16$, corresponding to $m=0.83$ and fit the data again.
The best fits in this case are:
$P_1=148.81\pm34.77$, $P_2=(2.85 \pm 0.68) \times 10^{-2}$, 
and $b=1.12\pm0.06$.
As expected, the evolution deduced during the optically thin case
is similar to that found in the FFA model.

We note that there are more complicated possibilities for the absorption,
but our data are not sufficient to test them.
In particular, \citet{wei90} found that the optically thick radio evolution
of the Type IIn SN 1986J could not be fit by a model with external FFA,
but could be fit by a model with internal FFA mixed with the synchrotron
emitting region.
Such a model is plausible for SNe IIn, because of the evidence for 
clumping in the CS medium, but we cannot test it here.

\begin{figure}
\includegraphics[angle=0,width=.49\textwidth]{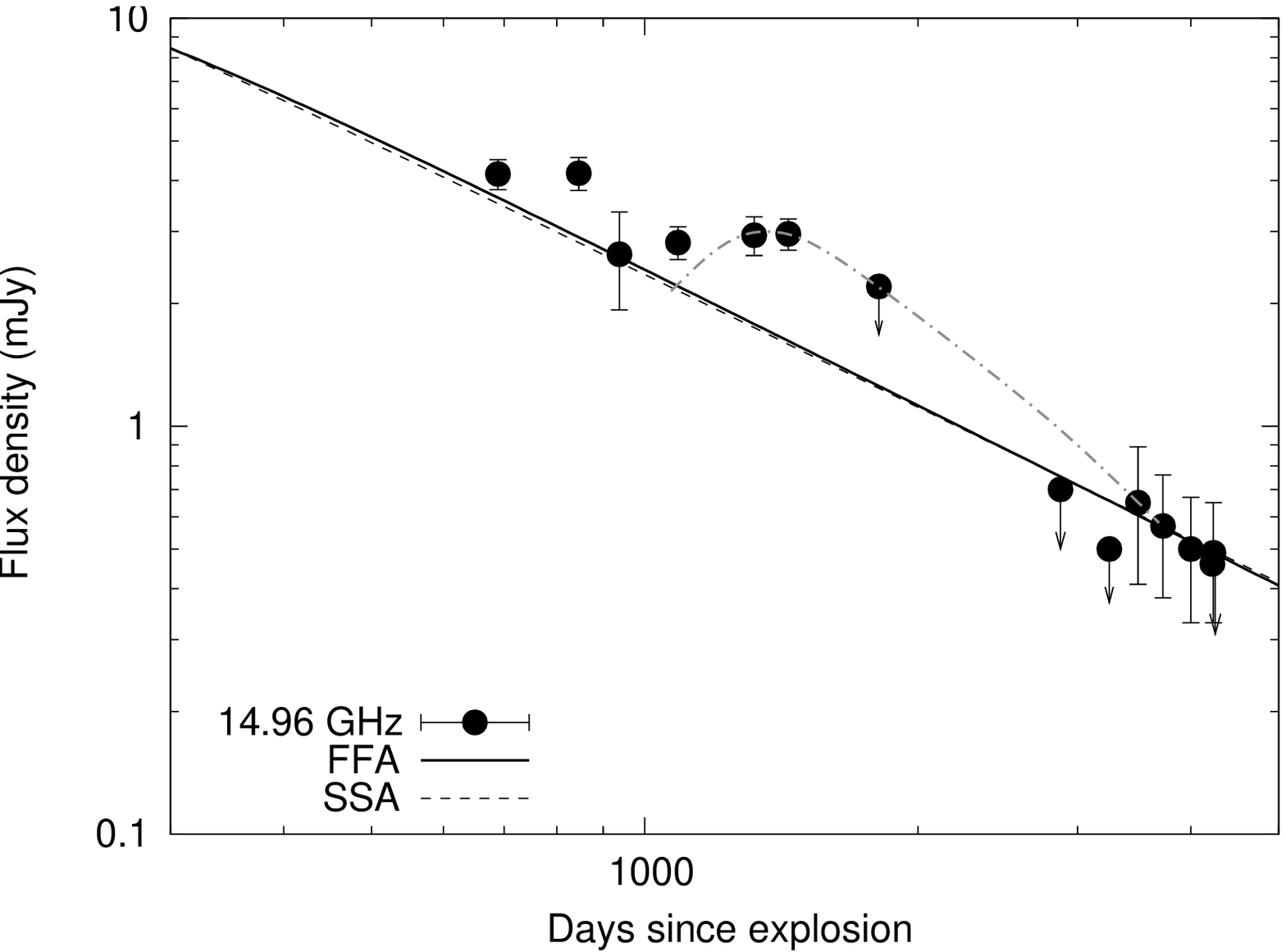}
\includegraphics[angle=0,width=.49\textwidth]{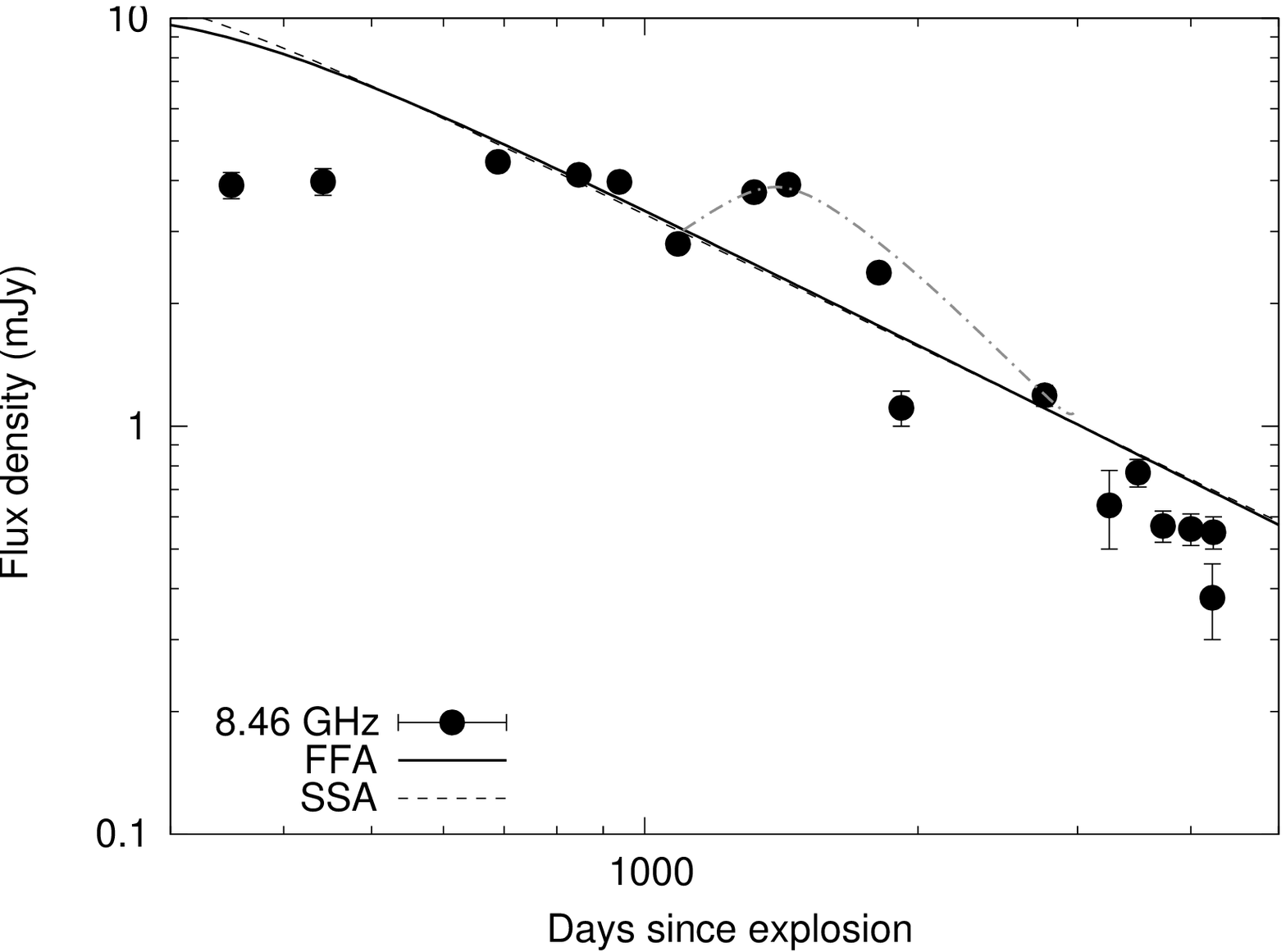}
\includegraphics[angle=0,width=.49\textwidth]{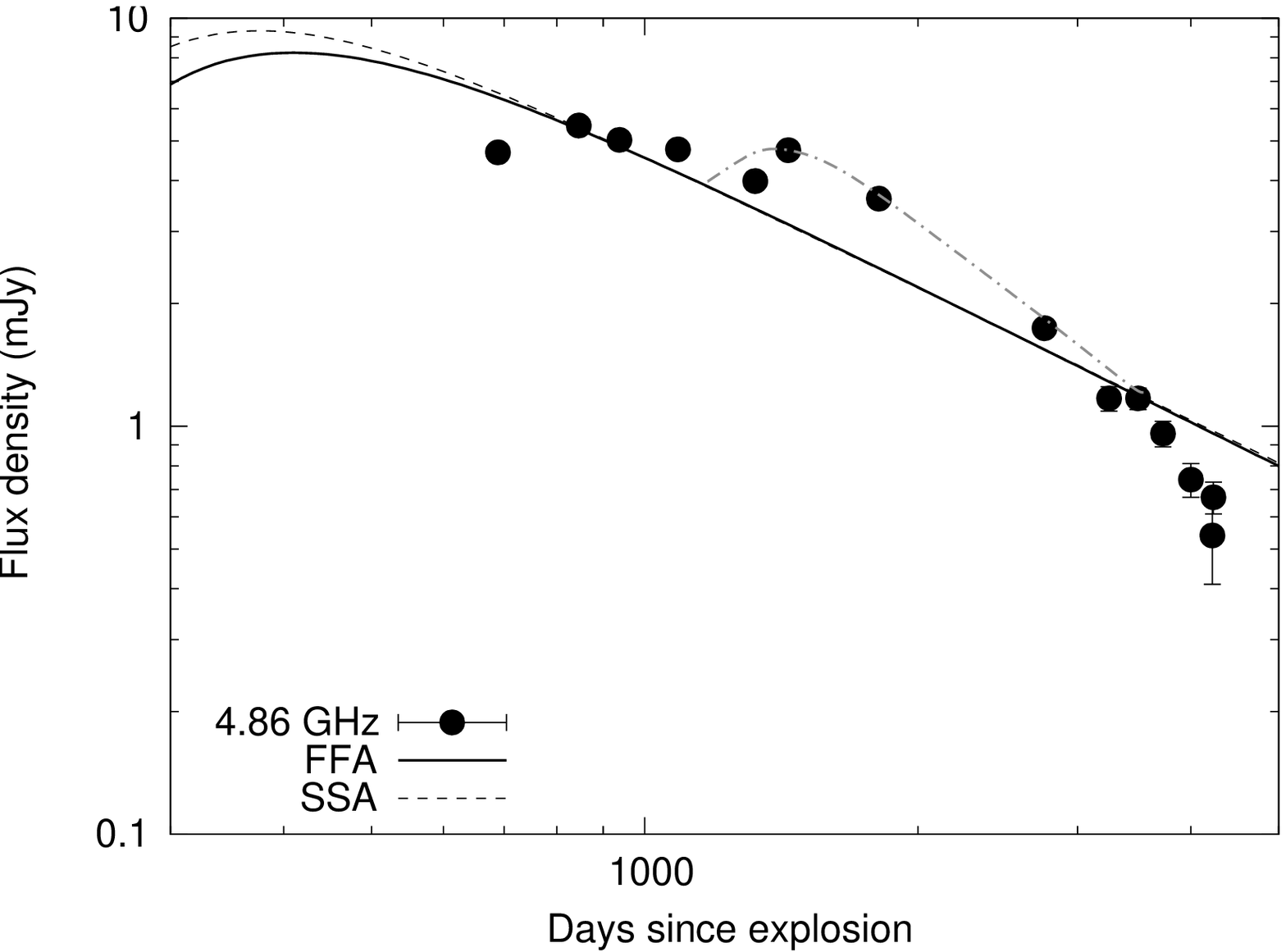}
\includegraphics[angle=0,width=.49\textwidth]{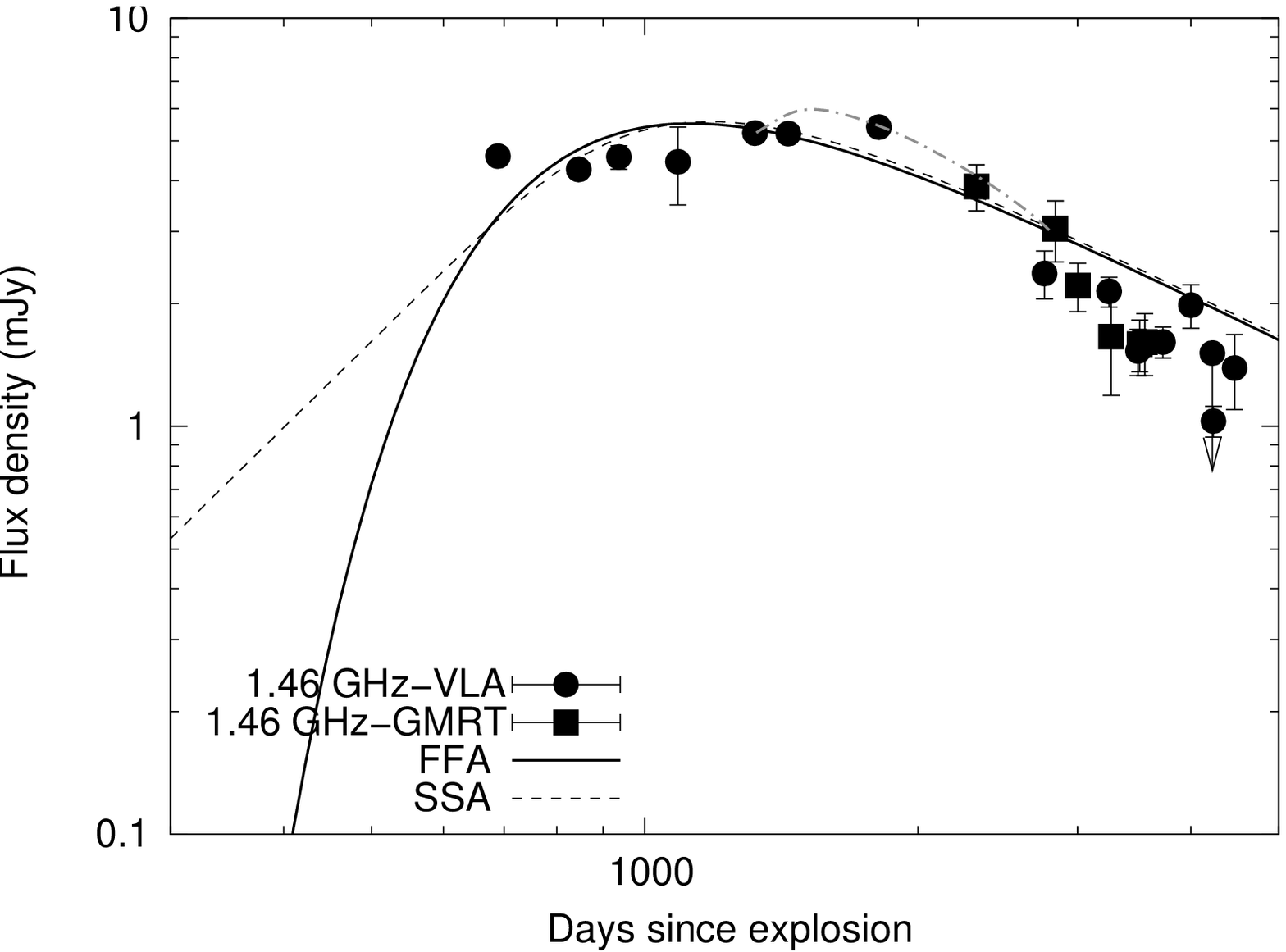}
\caption{Light curves of SN 1995N in the 2 cm, 3.6 cm, 6 cm and
1.46 cm bands. The FFA absorption model is shown in solid lines and
dashed lines represent the SSA absorption model. There
are hints of bumps in the 
 light curves 
soon after day 1200 in all the frequency bands (dashed-dotted lines),
which may be due to the density clumps in the CSM.}
\label{fig:lc}
\end{figure}

\begin{figure}
\includegraphics[angle=0,width=.47\textwidth]{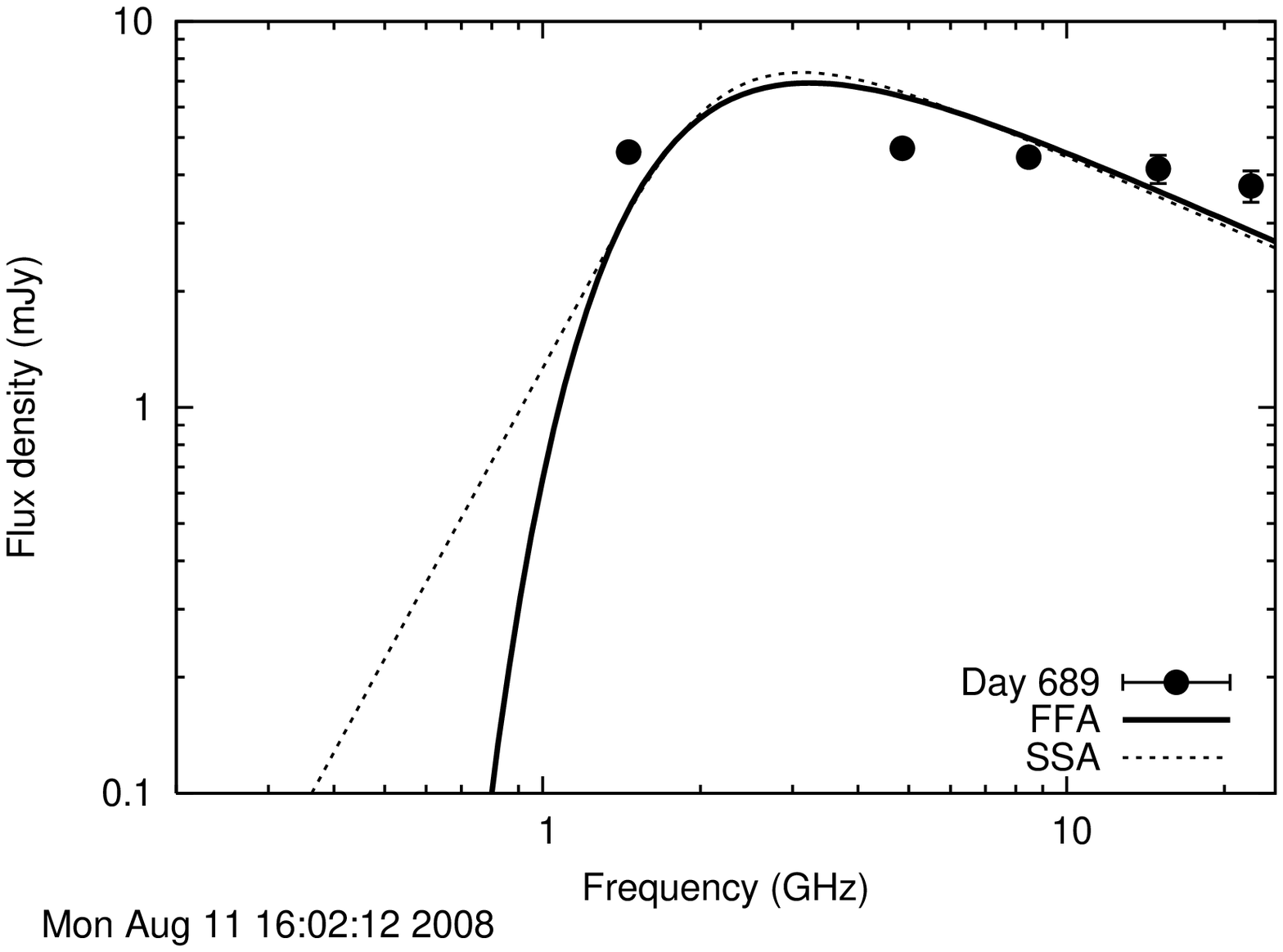}
\includegraphics[angle=0,width=.47\textwidth]{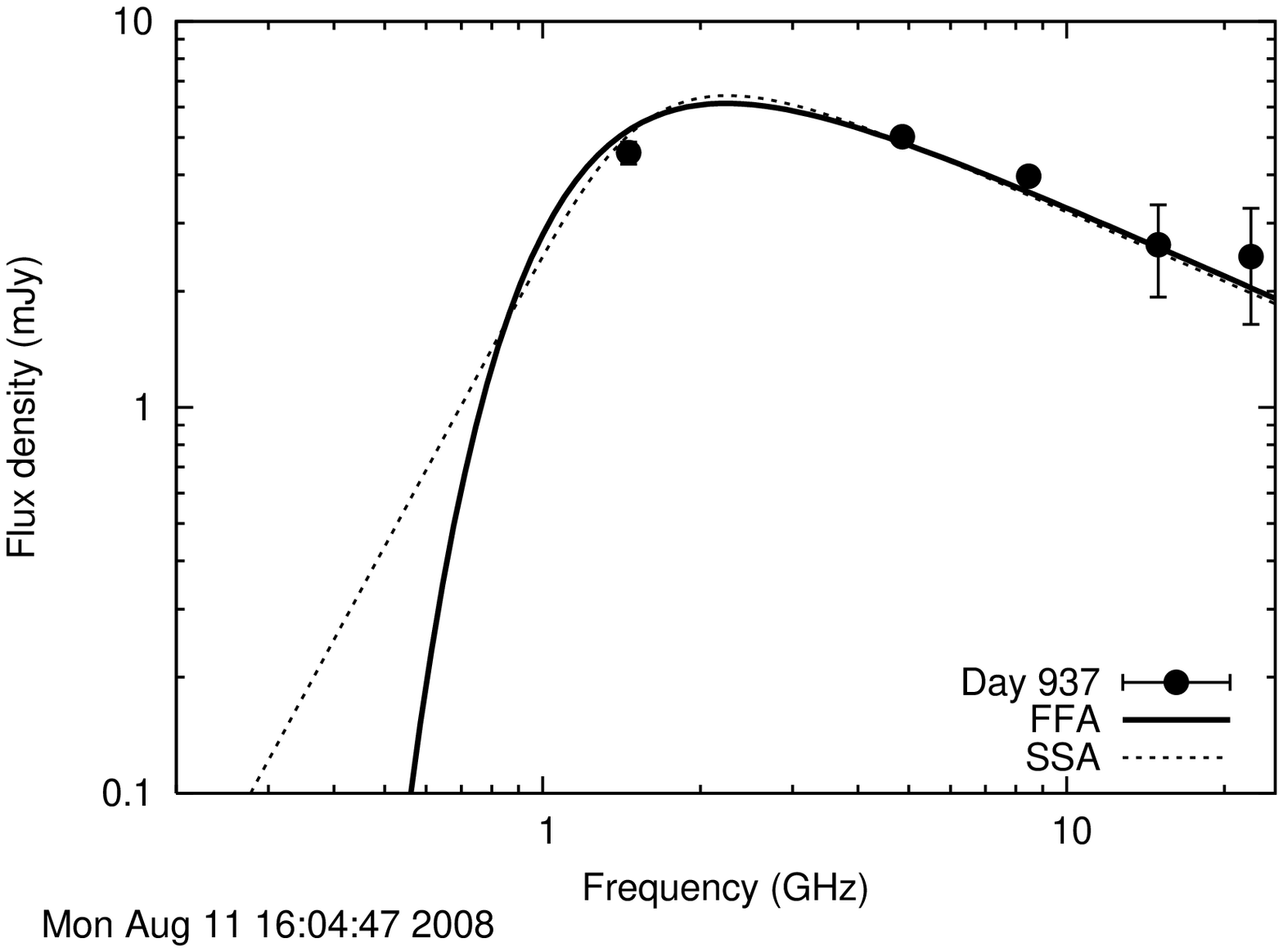}
\includegraphics[angle=0,width=.47\textwidth]{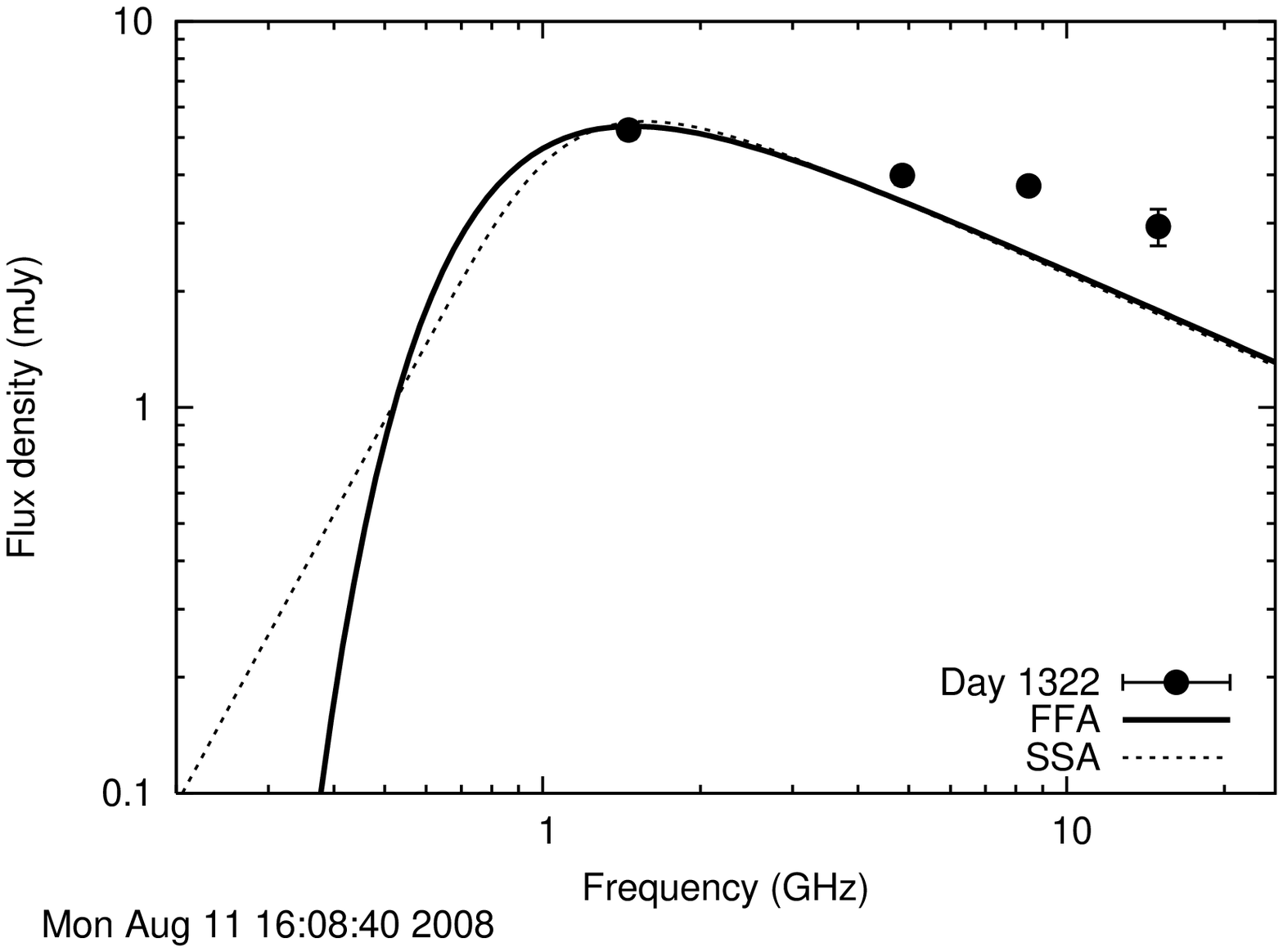}
\includegraphics[angle=0,width=.47\textwidth]{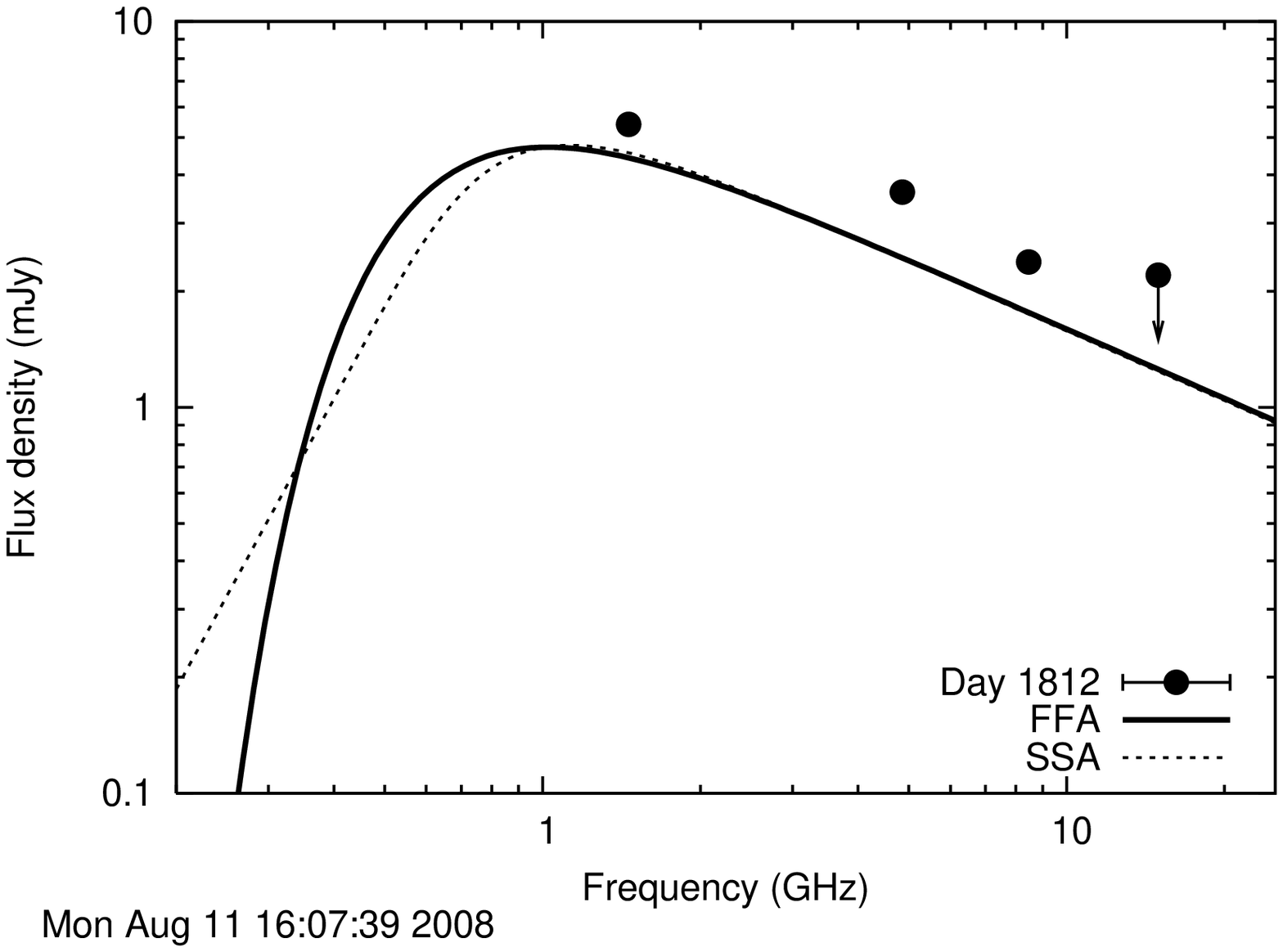}
\includegraphics[angle=0,width=.47\textwidth]{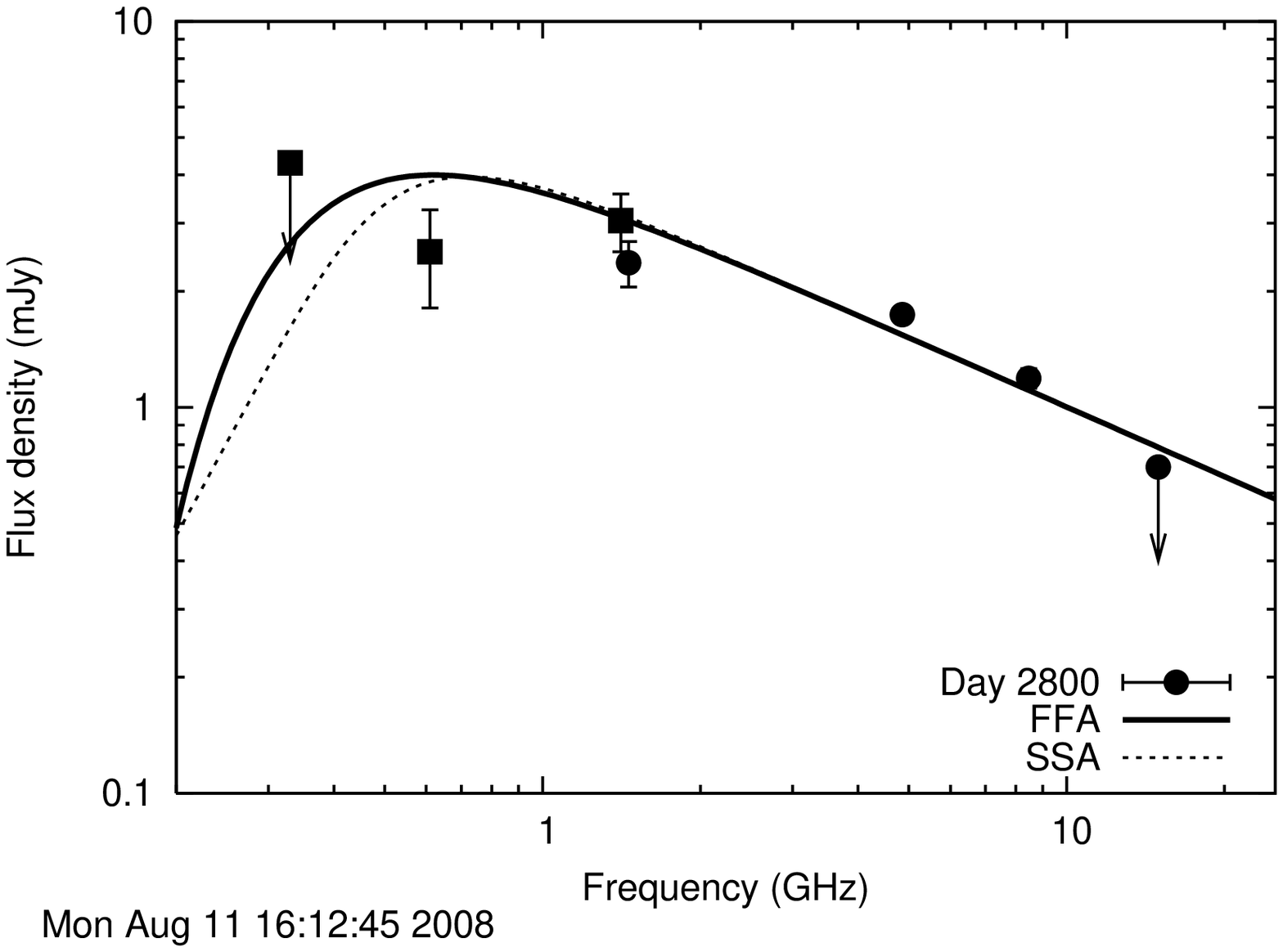}
\includegraphics[angle=0,width=.47\textwidth]{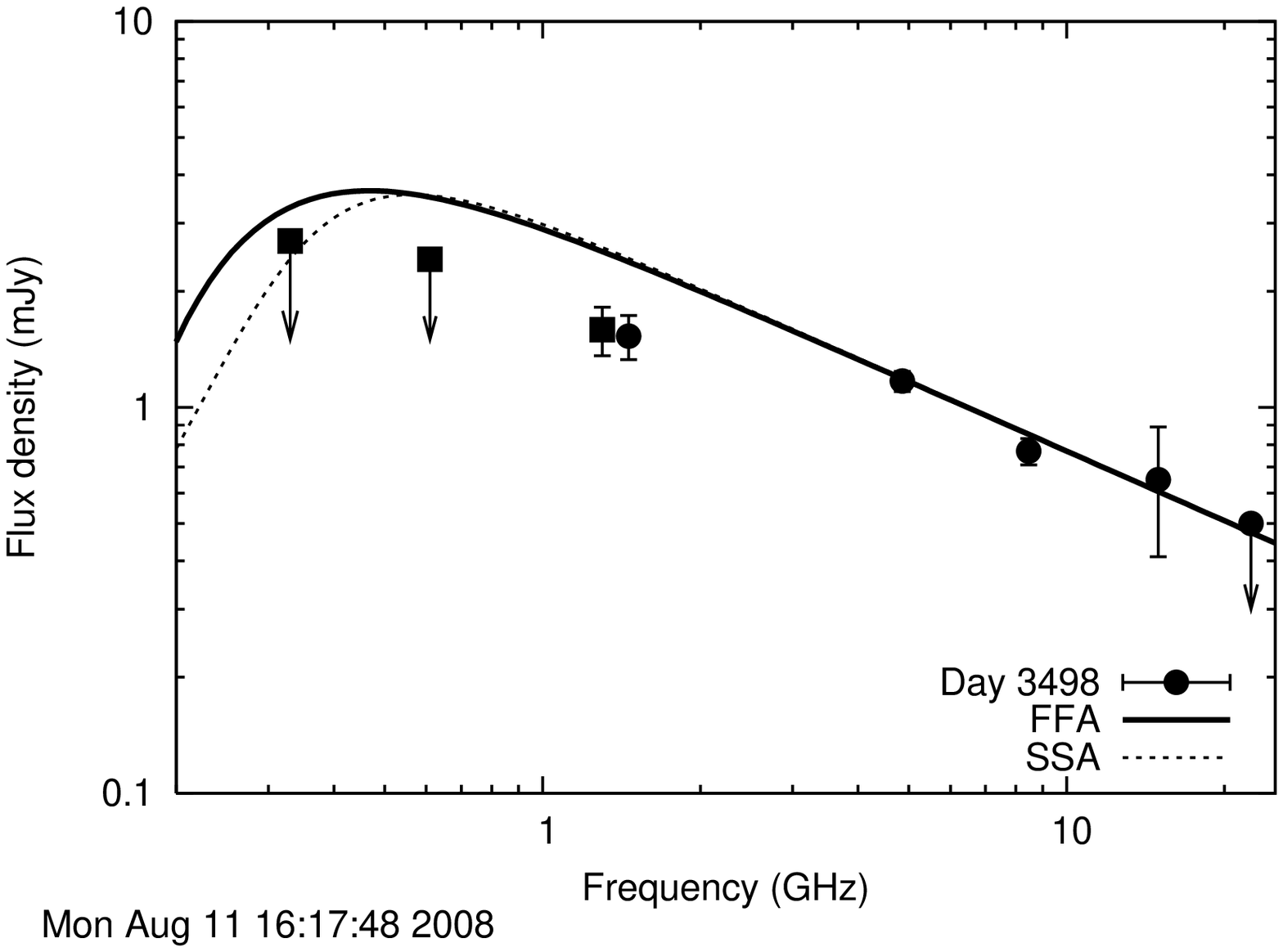}
\caption{
Near simultaneous spectra of SN 1995N on various days.
FFA absorption is shown in solid lines and
dashed lines represent the SSA absorption.
Here filled circles represent VLA data and filled squares represent GMRT data.
Please note that models do not fit the
spectra well on days when bumps are seen in the light curves, 
i.e. around day 1200.
}
\label{fig:spectra}
\end{figure}

Although both FFA and SSA give acceptable
fits due to the lack of data points in the optically thick regime,
the
SSA derived velocity and information using optical data suggest that FFA
is the most relevant
radio absorption process. We
produce light curves (see Fig. \ref{fig:lc}) and spectra (see
Fig. \ref{fig:spectra}) for both FFA and SSA processes.
In Fig. \ref{fig:lc}, there is a sudden rise in the
radio flux around day $\sim 1200$, plotted with grey dashed-dotted lines; 
this is evident in all the frequency
bands almost at the same time. 
This effect can also be see in the spectra on 
days 1440 and 1812 (Fig. \ref{fig:spectra}). 
We speculate that this feature is due to a density enhancement in the CSM. 
However, if the time decay $\beta$ is somewhat shallower
at the early epoch, and there is a steepening in  $\beta$ at later epochs,
this may also cause such a bump in the light curves. Such time
evolution in $\beta$ has been seen for
SN 1993J and a few other supernovae \citep{wei07}. 
It is difficult to distinguish between the two possibilities
in view of the sparse data.

\section{Physical Conditions}
\label{sec:interpretation}

SN 1995N was extensively observed in many wavebands of the electromagnetic
spectrum. We will utilize this information and
derive constraints on various physical parameters of SN 1995N.

\subsection{Mass loss rate}
\label{sec:massloss}

If FFA is the dominant absorption process, we
can  estimate the mass-loss rate of the progenitor star before it underwent the
explosion phase using the radio 
data. This can be written as \citep{wei86,van94}
\begin{equation}
\dot M= 2.11 \times 10^{-5}\left [ \frac{A_2}{10^{-2}}\right]^{0.5}\left 
[\frac{V_{ej}}{5,000\, {\rm km/s}}\right]^{1.5}\left 
[\frac{T_e}{20,000 \, {\rm K}}
\right]^{0.68} \left [\frac{u_{w}}{10 \, {\rm km/s}}\right]
\left[\frac{t}{1000\, {\rm day}}\right]^{3(1-m)/2}
{\rm M_\odot \, yr^{-1}},
\label{eq:massloss}
\end{equation}
where $V_{ej}$ is the ejecta velocity  % at time $t_i$ 
 and $T_e$ is the electron
temperature. $A_2$ is the best fit normalization constant in
the FFA  model (see Eq. \ref{eq:ffa}). 
The normalization values of $V_{ej}$ and $T_e$ are relevant for late time 
observations, say around the epoch of {\it Chandra} observations on
day $\sim 3550$ (see \S \ref{sec:density}).
Assuming $u_w\approx 10$ km s$^{-1}$ 
 gives a mass loss rate of $\dot M=(6.14 \pm 0.47) \times 10^{-5}$ 
${\rm M_\odot \, yr^{-1}}$; the error in mass loss rate is
obtained using the best fit error in $A_2$.  Uncertainties in the ejecta 
velocity, electron temperature and wind velocity introduce more
uncertainty in the mass loss rate estimation.
In addition, equation (\ref{eq:massloss}) is based on the assumption that the
absorbing wind gas is smoothly distributed.
However, \citet{fra02} found that the density of ionized gas obtained from line
diagnostics is higher than would be expected for smoothly distributed gas.
With clumping, there is less matter for the same amount of absorption and
the mass loss rate is reduced.

\subsection{Density and temperature}
\label{sec:density}

SN 1995N was  observed with XMM-Newton 
in July 2003 \citep{zam05} and with {\it Chandra}
in March 2004 \citep{cha05}. We can use this 
information to extract the density structure of the
SN ejecta around this time. 
It was estimated in  \citet{cha05} and \citet{zam05} that the 
total X-ray emission  for SN 1995N around the time of the XMM and
{\it Chandra} observations is 
$L_{X}=1.1 \times 10^{40}$ erg s$^{-1}$.
\citet{cha05} estimated that most of the 
X-rays are coming from the reverse shock, and that
there is a cooled shell present at the boundary of the 
reverse and forward shock with
a column density of the cooling shell to be $N_{cool}=(9.1 \pm 6.2)
\times 10^{20}$ cm$^{-2}$.
In the case of cooling, the total luminosity from the reverse shock
can be written as
\citep{che03}
\begin{equation}
L_{rev}=2 \times 10^{40}\,\frac{(n-3)(n-4)}{(n-2)^3} \left[\frac{\dot M}
{10^{-5}{\rm\, M_\odot \, yr^{-1}}}\right]  \left [\frac{u_{w}}{10 \,
 {\rm km \, s^{-1}}}\right]^{-1}
\left [\frac{
V_{ej}}{5,000\, {\rm km\,s^{-1}}}\right]^3 \, {\rm erg\, s^{-1}},
\label{eq:lumin}
\end{equation} 
where $n$ is the power law density index of the
 SN ejecta ($\rho_{ej}\propto r^{-n}$, $\rho_{ej}$ ejecta density of the
SN).
Using the mass loss rate values derived in the above section (\S 
\ref{sec:massloss}), Eq. \ref{eq:lumin} gives
the ejecta density index 
$n\approx 8.5$. 
This value of $n$ is  consistent with the one obtained 
from the radio evolution described in \S \ref{sec:absorption}.

We can write the expression for the column density of the cooled shell in 
terms of mass loss rate as \citep{che03}
\begin{equation}
N_{cool}=2\times 10^{20}\, (n-4)  \left[\frac{\dot M} {10^{-5}
{\rm M_\odot \, yr^{-1}}}\right]  
\left [\frac{u_{w}}{10 \,  {\rm km \, s^{-1}}}\right]^{-1}
\left [\frac{V_{ej}}{5,000\, {\rm km\,s^{-1}}}\right]^{-1}
\left[\frac{t}{1000\,{\rm days}}\right]^{-1} \,{\rm cm^{-2}}.
\label{eq:ncool}
\end{equation}
At the epoch of the {\it Chandra} observation, this gives 
$N_{cool}\approx 1.5 \times 10^{21}$ cm$^{-2}$, which is in agreement
with the value deduced from the X-ray observations 
\citep[$(9.1 \pm 6.2)
\times 10^{20}$ cm$^{-2}$,][]{cha05}.

Having an estimate for the mass loss rate and the density profile index,
one can estimate the temperatures in the shocked circumstellar shell and
the reverse shocked shell.
The temperature of the shocked CSM shell is \citep{che03}
\begin{equation}
T_{CS}=3.4\times 10^8 \frac{(n-3)^2}{(n-2)^2}\left
 [\frac{{V_{ej}}}{5,000\, {\rm km\,s^{-1}}}\right]^2\,{\rm K}
\label{eq:tempcs}
\end{equation}
and the reverse shock temperature is
\begin{equation}
T_{rev}=\frac{T_{CS}}{(n-3)^2}\, \rm K
\label{eq:temprev}
\end{equation}
The above equations give $T_{CS}=2.4 \times 10^8$ K and 
$T_{rev}=0.9 \times 10^7$ K. These estimates are derived assuming
cosmic abundances and equipartition between ions and electrons.
The equipartition time scale between electrons and ions is
$t_{eq}\approx290(T_e/10^9 \,\rm K)^{1.5}(n_e/10^7 \rm cm^{-3})^{-1}$ days.
\citet{cha05} estimate the electron density of the reverse shocked 
shell to be $n_{rev}\approx 2 \times 10^6$ cm$^{-3}$.   The density 
of the shocked CSM shell is thus $n_{cs}=2n_{rev}/(n-4)(n-3)=3.3 \times 10^5$
cm$^{-3}$. This gives the estimates for equipartition time 
scales in the reverse shocked and the forward shocked shells as
$1.3$ days and 1007.5 days, respectively. Hence, at the time of
{\it Chandra} and XMM observations, the epoch at which the above
parameters are derived, both the shocked 
shells have most likely attained electron-ion
equilibrium.

\section{Discussion and Conclusions}
\label{sec:conclusions}

A useful way to characterize the radio emission from supernovae is by
considering their peak radio luminosity and time of the peak;
the more luminous Type II supernovae peak at a later time, indicating
that they have a denser circumstellar medium \citep{che06}.
In such a plot, SN 1995N is close to the Type IIL SN 1979C and Type IIn
SN 1978K, but is less luminous 
that the Type IIn SNe 1986J and 1988Z \citep{wei90,wil02}.
However, it is more radio luminous than many of the SNe IIn observed
by \citet{van96}.

The optically thin radio evolution of SN 1995N is fairly well defined
and yields a radio spectral index $\alpha\approx 0.6$.
This value is in the range found for Type II SNe and is smaller than
the values found for Type Ib/c SNe \citep{wei02}.
The rate of decline is also similar to that found for other SNe II
and, in a simple model for the emission, indicates an expansion
factor for the emitting region $m=0.83$.
Assuming interaction with a steady wind, the implied value of the
power law index of the outer supernova density profile is $n\approx8$.
The same value of $n$ is consistent with the X-ray properties of
SN 1995N.

Although there are clear indications of early low frequency absorption
in the radio light curves of SN 1995N, there is little information on
the evolution during the optically thick phase so that the nature of
the absorption cannot be deduced from the radio light curves.
Based on the emission measure of the gas deduced from optical 
observations and the late  peak radio luminosity, we infer that
free-free absorption is the likely absorption mechanism.
The implied mass loss rate for the supernova progenitor is
$6 \times 10^{-5}\, {\rm M_\odot \, yr^{-1}}$, but this value has a number
of uncertainties.
The assumed velocity of the radio emitting region is $5000$ km s$^{-1}$;
however, there are some indications from the H$\alpha$ line that there are
higher velocities \citep{fra02}, which would lead to a higher value of $\dot M$.
The assumed wind velocity of 10 km s$^{-1}$ is lower than has been inferred
from the narrow H$\alpha$ lines observed in some SNe IIn, e.g., 90 km s$^{-1}$
in SN 1997ab \citep{sal98}, 160 km s$^{-1}$ in SN 1997eg \citep{sal02},
and 100 km s$^{-1}$ in SN 2002ic \citep{kot04}.
An order of magnitude increase in $u_w$ would increase $\dot M$ by an
order of magnitude.
Finally, the optical line emission suggests clumping of the ionized wind
gas \citep{fra02}, which would reduce the estimate of $\dot M$ found here.

Regardless of the uncertainty in $\dot M$, the values deduced for SNe IIn
from optical observations listed in \S~1 are significantly larger than
the value deduced for SN 1995N.
In some of those cases, the mass loss may be a short lived event before the
supernova.
The long time coverage of SN 1995N, especially at radio wavelengths, shows
that the circumstellar gas is extended in this case, out to radii $\ga 2\times
10^{17}$ cm.
The Type IIn supernovae are a heterogeneous group of objects.

\acknowledgments
We thank the VLA staff for making radio observations,
without which this work was not possible.
For the GMRT observations, we thank the staff of the GMRT
 which is run by the National Center 
for Radio Astrophysics of the Tata Institute 
of Fundamental Research (TIFR). We acknowledge
 the use of AIPS, which was developed
 by the staff of the National Radio Astronomical Observatory.
P.C. is a Jansky fellow at National Radio Astronomy
Observatory. The National Radio Astronomy Observatory is a
facility of the National
Science Foundation operated under cooperative agreement by Associated
Universities, Inc.
R.A.C. was supported in part by NSF grant AST-0807727.
At Tata Institute this research formed a part of the projects
10P-201 and 11P-904 of the Five Year Plans.

\clearpage

\begin{deluxetable}{cccccc}
%\tabletypesize{\scriptsize}
\tablecaption{Details of  phase calibration observations for SN 1995N
\label{tab:phase}}
\tablewidth{0pt}
\tablehead{
\colhead{Date of} &  \colhead{Telescope} & \colhead{Array} & 
\colhead{Frequency} & \colhead{Phase} & \colhead{Phase cal} \\
\colhead{observation} & 
\colhead{} & \colhead{config.} &
 \colhead{in GHz} & \colhead{calibrator} & \colhead{Flux (Jy)}
}
\startdata
1995 Jun 16.19 & VLA & DnA & 8.46 & 1507-168 & 2.31\\
1995 Sep 15.96 & VLA & BnA & 8.46 & 1507-168 & 2.30\\
1996 May 21.20 & VLA & CnD & 4.86 & 1507-168 & 2.25\\
1996 May 21.20 & VLA & CnD & 8.46 & 1507-168 & 1.95\\
1996 May 21.24 & VLA & CnD & 1.46 & 1507-168 & 2.83\\
1996 May 21.24 & VLA & CnD & 14.96 & 1507-168 & 1.75\\
1996 May 21.25 & VLA & CnD & 22.48 & 1507-168 & 1.59\\
1996 Oct 24.71 & VLA & A & 1.46 & 1504-166 & 2.52\\
1996 Oct 24.72 & VLA & A & 8.46 & 1504-166 & 2.12\\
1996 Oct 24.73 & VLA & A & 4.86 & 1504-166 & 2.32\\
1996 Oct 24.76 & VLA & A & 22.48& 1504-166 & 2.40\\
1996 Oct 24.76 & VLA & A & 14.96& 1504-166 & 2.27\\
1997 Jan 23.61 & VLA & BnA & 1.46 & 1504-166 & 2.46\\
1997 Jan 23.62 & VLA & BnA & 8.46 & 1504-166 & 2.21\\
1997 Jan 23.63 & VLA & BnA & 4.86 & 1504-166 & 2.38\\
1997 Jan 23.64 & VLA & BnA &14.96 & 1504-166 & 2.26\\
1997 Jan 23.64 & VLA & BnA &22.48 & 1504-166 & 2.30\\
1997 Jun 22.99  & VLA & BnC & 1.46 & 1504-166 & 2.71\\
1997 Jun 23.00 &VLA & BnC & 8.46 & 1504-166 & 2.11\\
1997 Jun 23.02  & VLA & BnC & 4.86 & 1504-166 & 2.33\\
1997 Jun 23.03  & VLA & BnC &14.96 & 1504-166 & 2.18\\
1997 Jun 23.04  & VLA & BnC &22.48 & 1504-166 & 2.04\\
1998 Feb 10.00  & VLA & DnA & 14.96 & 1504-166 & 2.91\\
1998 Feb 10.00  & VLA & DnA & 8.46 & 1504-166 & 2.61\\
1998 Feb 13.37  & VLA & DnA & 1.46 & 1504-166 & 2.68\\
1998 Feb 13.49  & VLA & DnA & 4.86 & 1504-166 & 2.57\\
1998 Jun 10.14  & VLA & BnA & 0.33 & 1504-166 & 1.63\\
1998 Jun 10.16  & VLA & BnA & 1.46 & 1504-166 & 3.07\\
1998 Jun 10.17  & VLA & BnA & 8.46 & 1504-166 & 3.13\\
1998 Jun 10.19  & VLA & BnA & 14.96 & 1504-166 & 3.84\\
1998 Jun 10.20 & VLA & BnA & 4.86 & 1504-166 & 2.80\\
1999 Jun 17.23 & VLA & DnA & 1.46 & 1504-166 & 2.77\\
1999 Jun 17.25 & VLA & DnA & 4.86 & 1504-166 & 2.78\\
1999 Jun 17.26 & VLA & DnA & 8.46 & 1504-166 & 2.68\\
1999 Jun 17.27 & VLA & DnA &14.96 & 1504-166 & 2.90\\
1999 Oct 01.89   & VLA & BnA & 8.46 & 1504-166 & 2.82\\
2000 Nov 08.00  & GMRT & $-$ & 1.42 & 1347+122 & 6.25\\
2001 Mar 25.00  & GMRT & $-$ & 0.61 & 1419+064 & 16.07\\
2002 Jan 19.65 & VLA & D & 1.46 & 1504-166 & 2.94 \\
2002 Jan 19.67  & VLA & D & 4.86 & 1504-166 & 2.54\\
2002 Jan 19.67  & VLA & D & 8.46 & 1504-166 & 2.13\\
2002 Apr 07.00   & GMRT & $-$ & 1.41 & 1445+099 & 2.40\\ 
2002 May 15.24   & VLA & A & 14.96 & 1504-166 & 1.70 \\
2002 May 19.00 & GMRT & $-$ & 0.61 & 1419+064 & 16.17 \\
2002 Aug 16.00 & GMRT & $-$ & 0.33 & 1419+064 & 29.65 \\
2002 Sep 16.00 & GMRT & $-$ & 0.61 & 1419+064 & 16.17 \\
2002 Sep 21.00 & GMRT & $-$ & 1.41 & 1351-148 & 1.12 \\
2003 May 26.30  & VLA & A & 1.46 & 1504-166 & 2.99 \\
2003 May 26.31  & VLA & A & 4.86 & 1504-166 & 2.34\\
2003 May 28.22  & VLA & A & 8.46 & 1504-166 & 2.03\\
2003 May 28.24   & VLA & A & 14.96 & 1504-166 & 1.64\\
2003 Jun 13.00  & GMRT & $-$ & 1.30 & 1432-180 & 0.81 \\
2003 Jun 16.00  & GMRT & $-$ & 0.61 & 1419+064 & 15.18 \\
2003 Jun 16.00  & GMRT & $-$ & 0.24 & 1419+064 & 44.58 \\
2004 Jan 29.48& VLA & BnC & 1.46 & 1504-166 & 2.99 \\
2004 Jan 29.50& VLA & BnC & 4.86 & 1504-166 & 2.37 \\
2004 Jan 29.51& VLA & BnC &14.96 & 1504-166 & 1.71 \\
2004 Jan 29.52& VLA & BnC & 8.46 & 1504-166 & 2.05 \\
2004 Feb 13.00  & GMRT & $-$ & 1.30 & 1432-180 & 1.01\\ 
2004 Mar 29.00  & GMRT & $-$ & 1.30 & 1432-180 & 1.00 \\
2004 Apr 01.00  & GMRT & $-$ & 0.33 & 1419+064 & 26.57 \\
2004 Apr 08.00  & GMRT & $-$ & 0.61 & 1419+064 & 19.71 \\
2004 Apr 08.00  & GMRT & $-$ & 0.24 & 1419+064 & 40.96 \\
2004 Sep 06.99  & VLA & A & 22.460 & 1512-090 & 4.96 \\
2004 Sep 12.94 & VLA & A & 1.46 & 1507-168 & 2.86 \\
2004 Sep 12.95 & VLA & A & 4.86 & 1507-168 & 2.52 \\
2004 Sep 12.96 & VLA & A & 8.46 & 1507-168 & 1.97 \\
2004 Sep 12.97 & VLA & A &14.96 & 1507-168 & 1.66 \\
2005 Jun 14.30 & VLA & BnC & 1.46 & 1507-168 & 2.93 \\
2005 Jun 14.31 & VLA & BnC & 4.86 & 1507-168 & 2.40\\
2005 Jun 14.32 & VLA & BnC & 8.46 & 1507-168 & 2.09\\
2005 Jun 14.33 & VLA & BnC &14.96 & 1507-168 & 1.75\\
2006 Jan 23.60   & VLA & D & 1.46 & 1507-168 & 2.77\\
2006 Jan 23.61  & VLA & D & 4.86 & 1507-168 & 2.24 \\
2006 Jan 23.63  & VLA & D & 8.46 & 1507-168 & 1.87 \\
2006 Jan 23.65  & VLA & D &14.96 & 1507-168 & 1.53\\
2006 Feb 02.59  & VLA & A & 4.86 & 1507-168 & 2.16\\
2006 Feb 02.60  & VLA & A & 1.46 & 1507-168 & 2.27\\
2006 Feb 02.62  & VLA & A & 8.46 & 1507-168 & 1.91\\
2006 Feb 02.63  & VLA & A &14.96 & 1507-168 & 0.77\\
2006 Sep 26.00  & VLA & BnC & 8.46 & 1507-168 & 1.75\\ 
2006 Sep 26.00  & VLA & BnC & 1.46 & 1507-168 & 2.62\\
2006 Sep 26.01 &  VLA & BnC & 4.86 & 1507-168 & 2.42\\
\enddata
\end{deluxetable}
\clearpage

\begin{deluxetable}{ccccccc}
%\tabletypesize{\scriptsize}
\tablecaption{Radio observations of the SN 1995N
\label{tab:radio}}
\tablewidth{0pt}
\tablehead{
\colhead{Date of} & \colhead{Days since} & \colhead{Telescope} & \colhead{Array} & 
\colhead{Frequency} & 
\colhead{RMS} & \colhead{Flux density}\\
\colhead{observation} & \colhead{explosion} & 
\colhead{} & \colhead{config.} & \colhead{in GHz} 
 & \colhead{$\mu$Jy} & \colhead{mJy}
}
\startdata
1995 Jun 16.19 & 350.19 &  VLA & DnA & 8.46 & 210 & $ 3.90 \pm 0.29 $\\
1995 Sep 15.96 & 441.96 & VLA & BnA & 8.46  & 220 &$ 3.98 \pm 0.30$ \\
1996 May 21.20 & 688.76 & VLA & CnD & 4.86  & 66 & $ 4.69\pm 0.12$\\
1996 May 21.20 & 688.76 & VLA & CnD & 8.46  & 58 & $ 4.45\pm 0.09$\\
1996 May 21.24 & 688.80 & VLA & CnD & 1.46  & 167 & $ 4.37 \pm 0.77 $\\
1996 May 21.24 & 689.02& VLA & CnD & 14.96 &  186 & $4.15 \pm 0.35$\\
1996 May 21.25 & 689.03 & VLA & CnD & 22.48&  249 & $ 3.75\pm 0.35$\\
1996 Oct 24.71 & 845.52 & VLA & A & 1.46  & 106 & $ 4.26\pm 0.19$\\
1996 Oct 24.72 & 845.53 & VLA & A & 8.46  & 75 & $ 4.13\pm 0.12$\\
1996 Oct 24.73 & 845.54 & VLA & A & 4.86  & 66 & $ 5.46\pm 0.12$\\
1996 Oct 24.76 & 845.57 & VLA & A & 22.48 & 328 & $ 3.55\pm 0.55$\\
1996 Oct 24.76 & 845.57 & VLA & A & 14.96 & 208 & $ 4.17\pm 0.39$\\
1997 Jan 23.61 & 936.42 & VLA & BnA & 1.46 & 178  & $ 4.57\pm 0.30$\\
1997 Jan 23.62 & 937.43 & VLA & BnA & 8.46 & 94 & $ 3.74\pm 0.16$\\
1997 Jan 23.63 & 937.44 & VLA & BnA & 4.86 & 105 & $5.03 \pm 0.18$\\
1997 Jan 23.64 & 937.45 & VLA & BnA &14.96 & 408 & $ 2.64\pm 0.71$\\
1997 Jan 23.64 & 937.45 & VLA & BnA &22.48 & 622 & $ 2.46\pm 0.82$\\
1997 Jun 22.99 & 1087.80 & VLA & BnC & 1.46 & 170 & $ 4.45\pm 0.96$\\
1997 Jun 23.00 & 1087.81&VLA & BnC & 8.46 &  81 & $ 2.80\pm 0.04$\\
1997 Jun 23.02 & 1087.83 & VLA & BnC & 4.86 & 55 & $4.77 \pm0.09 $\\
1997 Jun 23.03 & 1087.84 & VLA & BnC &14.96 & 155 & $2.82 \pm0.26 $\\
1997 Jun 23.04 & 1087.85 & VLA & BnC &22.48 & 320 & $1.95 \pm 0.63 $\\
1998 Feb 10.00 & 1320.00 & VLA & DnA & 14.96&  233 & $2.94 \pm 0.32$\\
1998 Feb 10.00 & 1320.00 & VLA & DnA & 8.46 & 37 & $ 3.75\pm 0.13$\\
1998 Feb 13.37 & 1323.22 & VLA & DnA & 1.46 & 82 & $ 5.23\pm0.27 $\\
1998 Feb 13.49 & 1323.35 & VLA & DnA & 4.86 & 71 & $ 3.99\pm0.21 $\\
1998 Jun 10.14 & 1439.99 & VLA & BnA & 0.33 & 12504 & $ <25.00 $\\
1998 Jun 10.16 & 1440.01 & VLA & BnA & 1.46 & 101 & $ 5.21\pm 0.17$\\
1998 Jun 10.17 & 1440.03 & VLA & BnA & 8.46 & 41 & $ 3.91\pm 0.07$\\
1998 Jun 10.19 & 1440.04 & VLA & BnA & 14.96& 160 &  $ 2.96 \pm 0.26$\\
1998 Jun 10.20 & 1440.05& VLA & BnA & 4.86 & 44 & $ 4.75\pm 0.08$\\
1999 Jun 17.23 & 1812.08& VLA & DnA & 1.46 & 58 & $ 5.41\pm 0.12$\\ 
1999 Jun 17.25 & 1812.10& VLA & DnA & 4.86 & 66 & $ 3.61\pm 0.11$\\ 
1999 Jun 17.26 & 1812.11& VLA & DnA & 8.46 & 60 & $ 2.58\pm 0.10$\\ 
1999 Jun 17.27 & 1812.12& VLA & DnA &14.96 & 195 & $ <2.2 $\\
1999 Oct 01.89 & 1918.74  & VLA & BnA & 8.46  & 46 & $1.11 \pm 0.11$\\
2000 Nov 08.00 & 2320.85 & GMRT & $-$ & 1.42 &  290 & $ 3.87\pm0.50 $\\
2001 Mar 25.00 & 2457.85 & GMRT & $-$ & 0.61 &  130 & $ 3.26\pm 0.27$\\
2002 Jan 19.65 & 2758.51& VLA & D & 1.46 & 91  & $ 2.37\pm 0.32$\\ 
2002 Jan 19.67 & 2758.53 & VLA & D & 4.86 & 46 & $1.74 \pm 0.08$\\ 
2002 Jan 19.67 & 2758.53 & VLA & D & 8.46 & 39 & $1.19 \pm0.07 $\\
2002 Apr 07.00 & 2835.86  & GMRT & $-$ & 1.41  & 310 & $ 3.05\pm 0.52$\\
2002 May 15.24 & 2874.09  & VLA & A & 14.96 & 232 & $<0.70 $\\
2002 May 19.00 &2877.86 & GMRT & $-$ & 0.61  & 130 & $ 2.17\pm0.27 $\\
2002 Aug 16.00 &2966.86 & GMRT & $-$ & 0.33  & 780 & $ <4.30 $\\
2002 Sep 16.00 &2997.86 & GMRT & $-$ & 0.61  & 390 & $ 2.53\pm 0.72$\\
2002 Sep 21.00 &3002.86 & GMRT & $-$ & 1.41 & 100 & $2.21 \pm 0.30$\\
2003 May 26.30 & 3250.15 & VLA & A & 1.46  & 97 & $ 2.14\pm 0.18$\\
2003 May 26.31 & 3250.16 & VLA & A & 4.86  & 51 & $ 1.17\pm 0.08$\\
2003 May 28.22 & 3252.09 & VLA & A & 8.46  & 44 & $ 0.64\pm 0.14$\\
2003 May 28.24 & 3252.11  & VLA & A & 14.96 &  166 & $ <0.50 $\\
2003 Jun 13.00 & 3267.87 & GMRT & $-$ & 1.30 & 380 & $ 1.66\pm 0.47$\\
2003 Jun 16.00 & 3270.87 & GMRT & $-$ & 0.61  & 190 & $ 2.29\pm 0.39$\\
2003 Jun 16.00 & 3270.87 & GMRT & $-$ & 0.24  & 3030 & $ <5.00 $\\
2004 Jan 29.48& 3498.23 & VLA & BnC & 1.46  & 115 & $ 1.53\pm0.20 $\\  
2004 Jan 29.50& 3498.25 & VLA & BnC & 4.86  & 43 & $ 1.17\pm0.07 $\\  
2004 Jan 29.51& 3498.26 & VLA & BnC &14.96  & 134 & $ 0.65\pm0.24 $\\
2004 Jan 29.52& 3498.27 & VLA & BnC & 8.46  & 36 & $ 0.77\pm0.06 $\\  
2004 Feb 13.00 & 3512.75 & GMRT & $-$ & 1.30 & 210 & $1.59 \pm0.23 $\\
2004 Mar 29.00 & 3557.75 & GMRT & $-$ & 1.30 & 140 & $ 1.61\pm 0.28$\\
2004 Apr 01.00 & 3560.75 & GMRT & $-$ & 0.33  & 840 & $ <2.4 $\\
2004 Apr 08.00 & 3567.75 & GMRT & $-$ & 0.61  & 140 & $ <2.42 $\\
2004 Apr 08.00 & 3567.75 & GMRT & $-$ & 0.24  & 1740 & $ <3.50 $\\
2004 Sep 06.99 & 3720.99 & VLA & A & 22.460  & 187 & $ <0.50 $\\
2004 Sep 12.94 & 3726.94& VLA & A & 1.46  & 60 & $ 1.61\pm 0.14$\\  
2004 Sep 12.95 & 3726.95& VLA & A & 4.86  & 52 & $ 0.96\pm 0.07$\\  
2004 Sep 12.96 & 3726.96& VLA & A & 8.46  & 42 & $ 0.57\pm 0.05$\\  
2004 Sep 12.97 & 3726.97& VLA & A &14.96  & 189 & $ 0.57\pm 0.19$\\ 
2005 Jun 14.30 & 4001.30& VLA & BnC & 1.46  & 184 & $ 1.98\pm 0.24$\\ 
2005 Jun 14.31 & 4001.31& VLA & BnC & 4.86  & 56 & $ 0.74\pm 0.07$\\ 
2005 Jun 14.32 & 4001.32& VLA & BnC & 8.46  & 44 & $ 0.56\pm 0.05$\\ 
2005 Jun 14.33 & 4001.33& VLA & BnC &14.96  & 166 &$ 0.50\pm 0.17$ \\
2006 Jan 23.60 & 4225.08  & VLA & D & 1.46  & 551 & $ <1.51 $\\ 
2006 Jan 23.61 & 4225.09 & VLA & D & 4.86 & 80 & $ 0.54\pm 0.13$\\ 
2006 Jan 23.63 & 4225.11 & VLA & D & 8.46 & 55 & $ 0.38\pm 0.08$\\ 
2006 Jan 23.65 & 4225.13 & VLA & D &14.96 & 152 &$ <0.46 $ \\
2006 Feb 02.59 & 4234.59 & VLA & A & 4.86 & 47 & $ 0.67\pm 0.06$\\
2006 Feb 02.60 & 4234.60 & VLA & A & 1.46 & 52 & $ 1.03\pm 0.09$\\
2006 Feb 02.62 & 4234.62 & VLA & A & 8.46 & 40 & $ 0.55\pm0.05 $\\
2006 Feb 02.63 & 4234.63 & VLA & A &14.96 & 164 & $ 0.49\pm 0.16$\\
2006 Sep 26.00 & 4470.00 & VLA & BnC & 8.46 & 40 & $ <0.32 $\\ 
2006 Sep 26.00 & 4470.00 & VLA & BnC & 1.46 & 190 & $ 1.39 \pm 0.29 $\\ 
2006 Sep 26.01 & 4470.01 & VLA & BnC & 4.86 & 83 & $ 0.25 \pm 0.08 $\\ 
\enddata
\end{deluxetable}

\clearpage

\clearpage

\clearpage

\clearpage

\end{document}